\documentclass[fleqn,10pt]{wlscirep}
\usepackage[utf8]{inputenc}
\usepackage[T1]{fontenc}
\usepackage{amsmath}
\usepackage{marvosym}
\usepackage{graphicx,subfigure}         
\usepackage{amssymb,stmaryrd}   
\usepackage{color}
\usepackage{wrapfig}
\usepackage{mathtools}
\usepackage{tikz}  
\usetikzlibrary{arrows,shapes,trees,positioning}  
\usepackage{comment} 
\usepackage{units}
\usepackage{appendix}
\usepackage{hyperref}
\usepackage{textcomp}
\usepackage{ragged2e}
\usepackage[normalem]{ulem}

\DeclareUnicodeCharacter{2212}{\ensuremath{-}}
\DeclareUnicodeCharacter{03C0}{$\pi$}
\title{Direct evidence of terahertz emission arising from anomalous Hall effect
}

\author[1]{Venkatesh Mottamchetty}
\author[2]{Parul Rani}
\author[1]{Rimantas Brucas}
\author[1,2]{Anders Rydberg}
\author[1,*]{Peter Svedlindh}
\author[1,2,**]{Rahul Gupta}
\affil[1]{Department of Materials Science and Engineering, Uppsala University, Box 35, SE-751 03 Uppsala, Sweden}
\affil[2]{Department of Physics and Astronomy, Uppsala University, Box 516, SE-75120 Uppsala, Sweden}

\affil[**]{rahul.gupta@angstrom.uu.se}
\affil[*]{peter.svedlindh@angstrom.uu.se}

\begin{abstract}
A detailed understanding of the different mechanisms being responsible for terahertz (THz) emission in ferromagnetic (FM) materials will aid in designing  efficient THz emitters. In this report, we present direct evidence of THz emission from single layer Co$_{0.4}$Fe$_{0.4}$B$_{0.2}$ (CoFeB) FM thin films. The dominant mechanism being responsible for the THz emission is the anomalous Hall effect (AHE), which is an effect of a net backflow current in the FM layer created by the spin polarized current reflected at the interfaces of the FM layer. The THz emission from the AHE-based CoFeB emitter is optimized by varying its thickness, orientation, and pump fluence of the laser beam. Results from electrical transport measurements show that skew scattering of charge carriers is responsible for the THz emission in the CoFeB AHE-based THz emitter.
\end{abstract}
\begin{document}

\flushbottom
\maketitle

\section*{Introduction}
The region of the electromagnetic spectrum that lies between near microwave and far-infrared radiation is the so-called Terahertz (THz) radiation or THz gap, \textit{i.e.} typically frequencies between 100 GHz and 30 THz. Terahertz radiation finds applications in various fields, such as medicine, security, etc. \cite{valuvsis2021roadmap,federici2010review}.~Photoconductive switching, optical rectification, transient photo-current in air plasma, and difference frequency generation constitute techniques that are employed for the generation of THz radiation \cite{smith1988subpicosecond,venkatesh2014optical,lepeshov2017enhancement,jazbinsek2019organic,vediyappan2019,yang2007large,kim2007terahertz,yan2017high,PhysRevA.96.053402,bagley2018laser,fulop2020laser}.~Moreover, THz emission from magnetic materials, utilizing the spin degree of freedom, has recently gained popularity as a promising framework for the generation of broadband radiation without any phonon absorption gaps and with intensity comparable to the standard zinc telluride THz source \cite{seifert2016efficient,gupta2021co2feal}. 

There are several possible mechanisms that can explain THz generation in spin-based systems. Beaurepaire \textit{et al.} \cite{Beaurepaire_1996} discovered ultrafast demagnetization (UDM) in 1996, showing that a ferromagnetic (FM) Ni film when demagnetized on a subpicosecond time scale by a femtosecond (fs) laser pulse excitation  generates THz radiation \cite{zhang2020ultrafast}. 
The THz radiation is in this case proportional to the second time-derivative of the magnetization \cite{beaurepaire2004coherent} and shows a linear dependence on the thickness of the FM layer. Recently, Kampfrath \textit{et al.} \cite{kampfrath2013terahertz,seifert2016efficient} discovered an alternative mechanism for THz generation, which utilizes the inverse spin Hall effect (iSHE) or inverse Rasbha Edelstein effect (iREE). Here, the generation mechanism requires a magnetic heterostructure consisting of an FM layer and a non-magnet (NM) layer that possesses a high spin-to-charge (S2C) conversion efficiency. In this mechanism, the amplitude of the THz emission is highly dependent on the S2C conversion efficiency. 
Recently, it has been shown that THz emitters can be designed using a single FM layer, which utilizes the anomalous Hall effect (AHE) phenomenon  \cite{yang2018anomalous,PhysRevApplied.12.054027,PhysRevB.104.064419,PhysRevB.90.214410}.~On the one hand, the UDM mechanism relies on bulk properties of a single FM layer, while on the other hand the AHE mechanism corresponds to a combined effect of interface and bulk properties, which will be further discussed below. 

A fs laser pulse when incident on a FM layer will excite hot electrons in the FM layer. The system achieves equilibrium  through electron-electron, electron-phonon and electron-magnon interactions. Before attaining the equilibrium with respect to  electron-electron interactions, the hot  electrons acquire a velocity of order  $10^6$ m/s and move within the FM layer in a super-diffusive manner  \cite{PhysRevLett.105.027203,PhysRevB.86.024404}. When reaching the FM/dielectric interfaces as indicated in Fig. 1,  electrons reflect back from the interfaces to form a net backflow  current ($j_{bf}$) along the film thickness direction \cite{PhysRevB.86.024404}. In the presence of the AHE, $j_{bf}$ is converted to a  transient  current ($j_t$) defined as $j_t = \theta_{AHE} (m \times j_{bf}$), where $\theta_{AHE}$ and $m$ are the anomalous Hall angle and the magnetization direction, respectively. The net backflow current depends on the dielectric properties of the interfaces, their roughness and  properties of the FM layer, such as $\theta_{AHE}$ and $m$. 

\begin{figure}
        \centering
    \includegraphics[width=8cm]{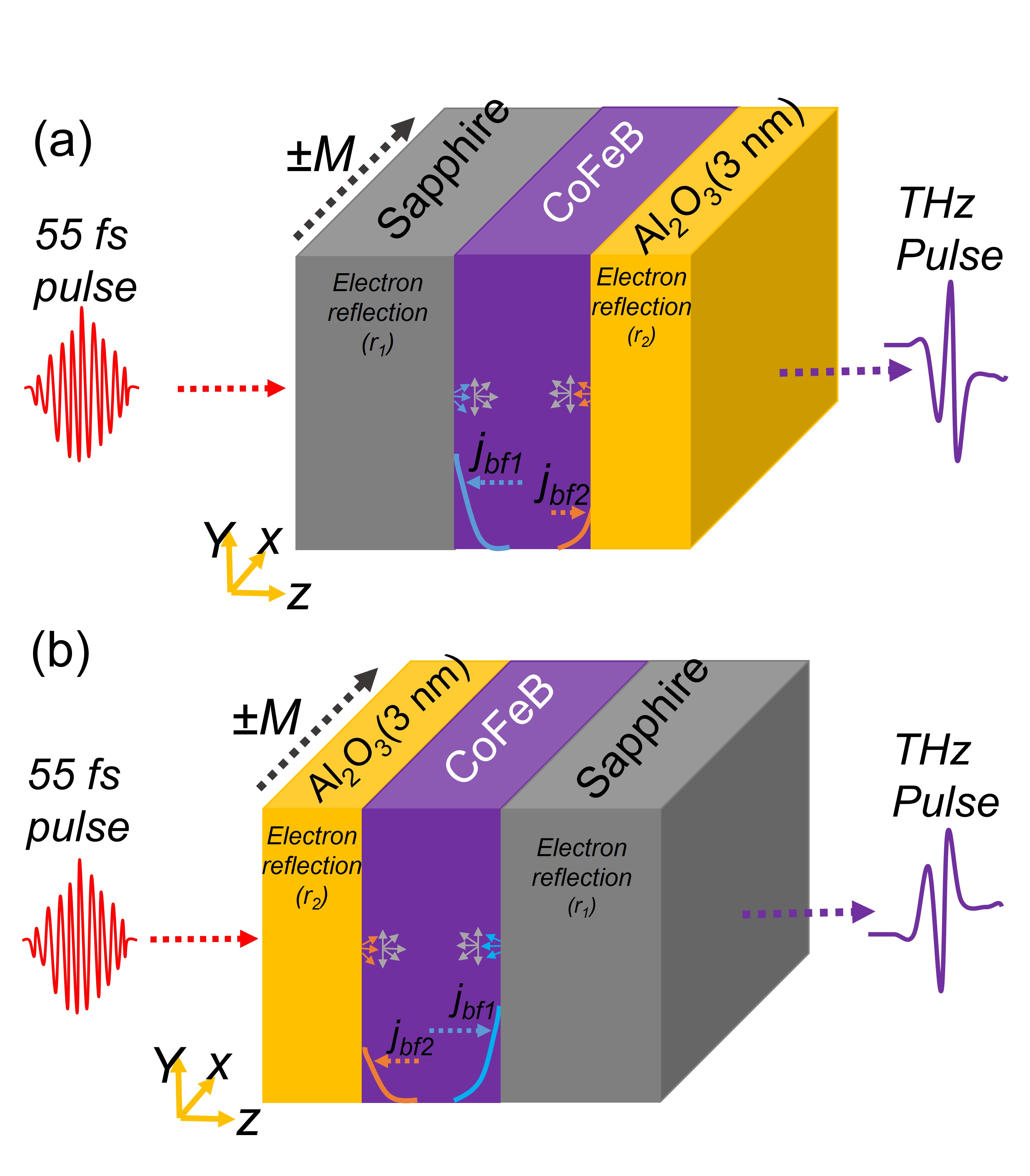}
        \caption{Schematic of THz emission from the AHE-based CoFeB emitter. (a) pumping from the substrate side and (b) pumping from capping side. j$_{bf1(2)}$ and r$_{1(2)}$ represent the backflow currents and electron reflection coefficients, respectively,  at the interface. Here, 1 and 2 correspond to the substrate/CoFeB  and CoFeB/capping interfaces, respectively.}
    \label{fig:THzschematic}
\end{figure}

Recently, Zhang \textit{et al.}~\cite{PhysRevApplied.12.054027} reported the THz emission in FM layers with different $\theta_{AHE}$, where (Fe$_{0.8}$Mn$_{0.2}$)$_{0.67}$Pt$_{0.33}$ (FeMnPt) with $\theta_{AHE}$=0.0269 was found to exhibit the largest THz emission. However, in the same study, the THz amplitude was found to be much smaller for Co$_{0.2}$Fe$_{0.6}$B$_{0.2}$ even though the anomalous Hall angle for Co$_{0.4}$Fe$_{0.4}$B$_{0.2}$ is of the same order of magnitude ($\theta_{AHE}^{CoFeB}\sim$ -0.023) as that of FeMnPt \cite{PhysRevB.90.214410}.
Moreover, NiFe was found to exhibit larger THz emission than Co$_{0.2}$Fe$_{0.6}$B$_{0.2}$, which is in contradiction with results reported by Seifert \textit{et al.} \cite{seifert2016efficient}. Note that the latter study utilized the iSHE mechanism for THz generation. Liu \textit{et al.} \cite{PhysRevB.104.064419} managed to distinguish  and separate the different  contributions to the  THz radiation, where they demonstrated that the UDM mechanism (magnetic dipole) is linearly  dependent on the FM layer thickness while the  AHE mechanism is not. In addition, the composition of the FM layer  defines the magnitude of $\theta_{AHE}$, which is a major factor in AHE-based THZ emitters  \cite{yang2018anomalous}. However, a systematic study providing direct evidence of the relationship between the anomalous Hall angle and the terahertz emission in single FM layers is still missing.

 In this study, we used  Co$_{0.4}$Fe$_{0.4}$B$_{0.2}$ (CoFeB) as the AHE layer to reveal the impact of the AHE on the THz emission.  We provide direct evidence for the relationship between the AHE and the THz emission in CoFeB, by presenting results from systematic studies of the effect of the CoFeB  thickness on the anomalous Hall resistivity, anomalous Hall angle and the anomalous Hall coefficient. In addition, we also reveal the mechanism behind the AHE in CoFeB. 

\begin{figure}
    \centering
    \includegraphics[width=12cm]{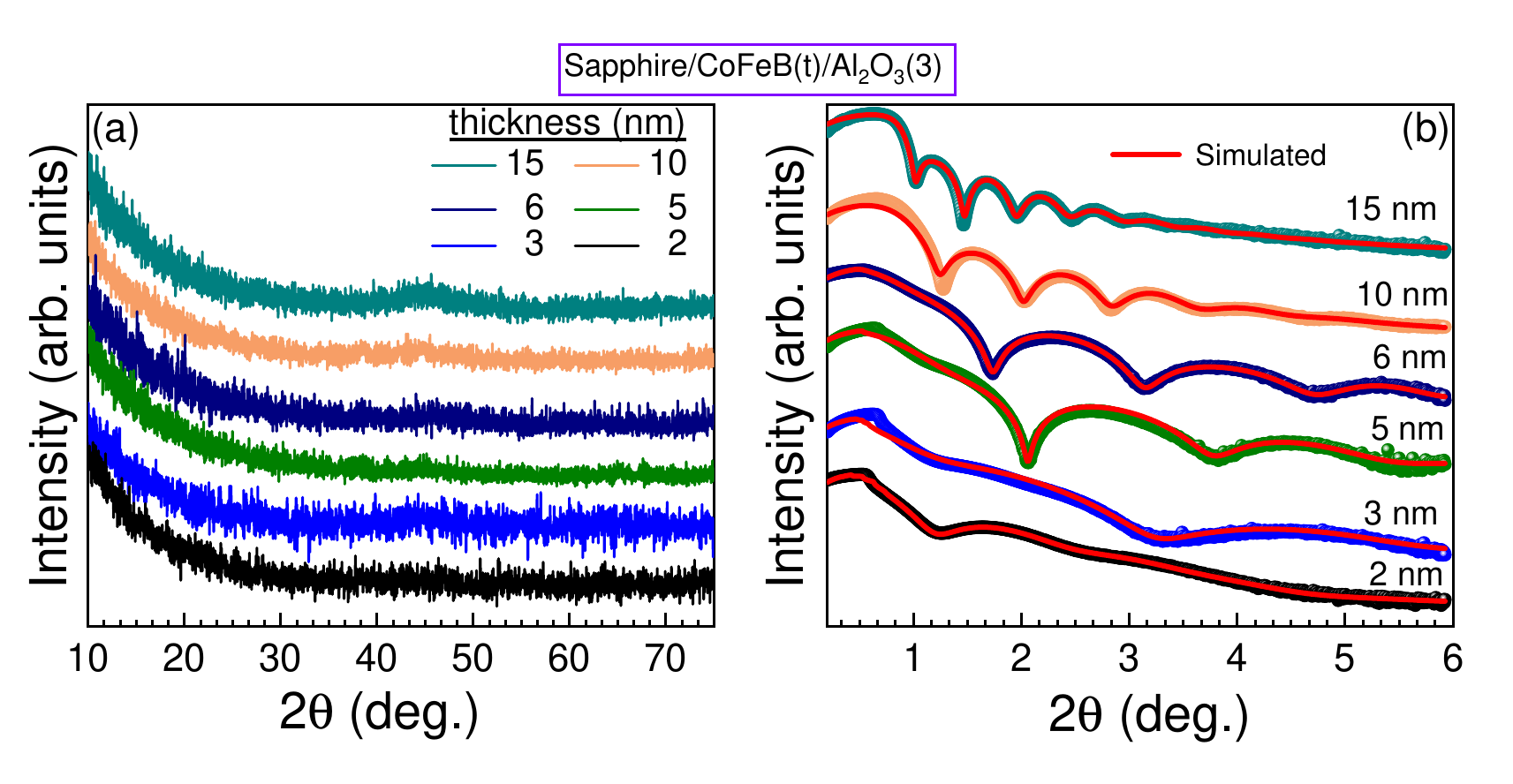}
        \caption{(a) XRD spectra for our emitters with different nominal thicknesses of CoFeB, and (b) XRR spectra for our emitters with different nominal thicknesses of CoFeB. Colored symbols and red lines correspond to experimental observed and simulated data, respectively.}
    \label{fig:XRD_XRR}
\end{figure}

\begin{figure}
    \centering
    \includegraphics[width=8cm]{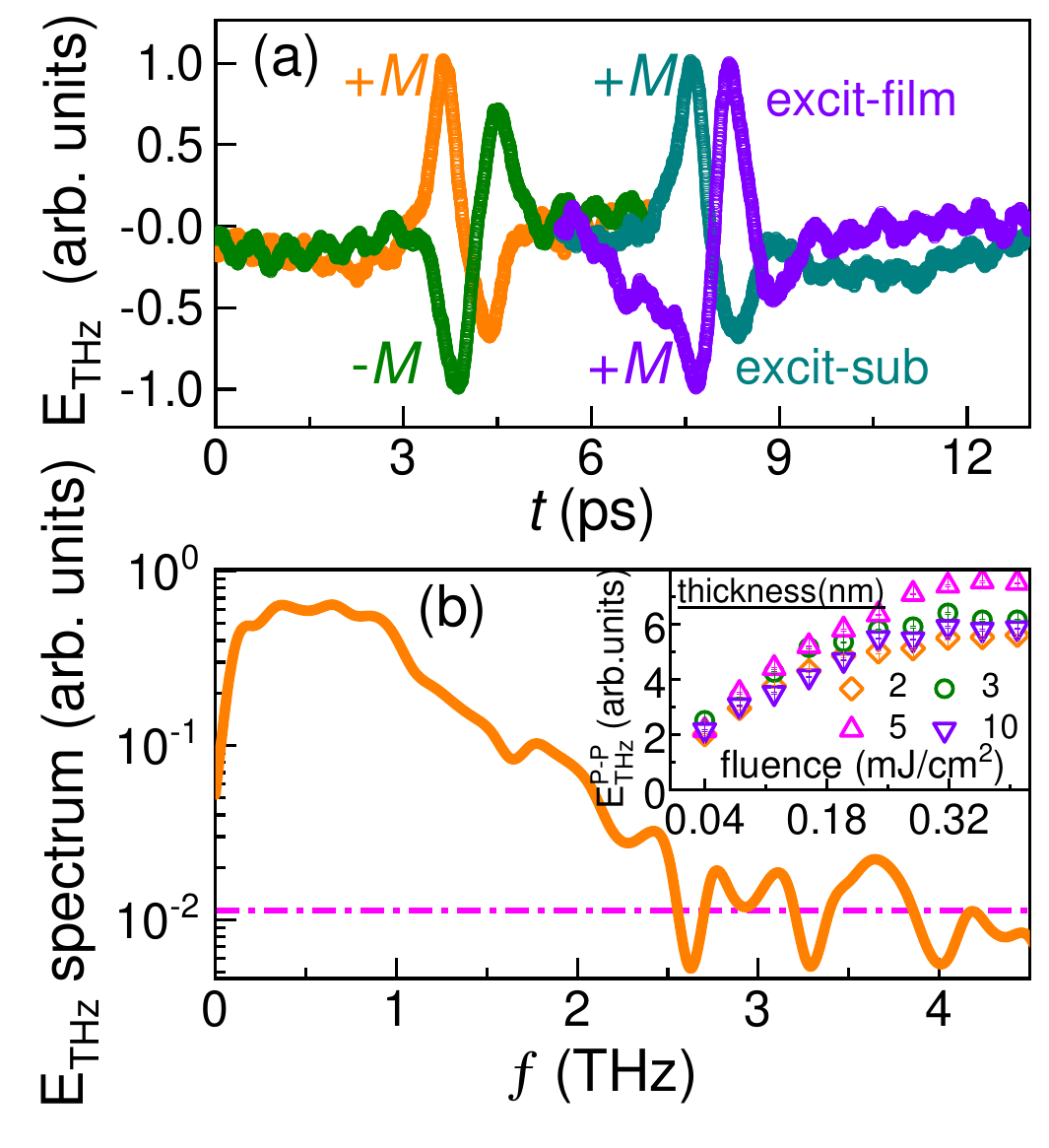}
        \caption{(a) Peaks at $\sim$4 ps: THz waveform in time-domain ($t$) for substrate/CoFeB(5 nm)/cap emitter with different magnetization directions ($\pm M$) when pumping from substrate side. Peaks at $\sim$8 ps: THz waveform pumping from the capping layer  and substrate sides while keeping the same magnetization direction. Time scale is shifted for better visualization of both cases. (b) Fast Fourier transform of THz electric field and the inset shows THz peak-to-peak amplitude for substrate/CoFeB($t_{CoFeB}$ nm)/cap emitters as a function of pump fluence.
        } \label{fig:THzemission}
\end{figure}

\section*{Results and discussion}

Thin films of CoFeB with different thicknesses ($t_{CoFeB}$) were deposited on Al$_2$O$_3$ substrates at room temperature using a DC magnetron sputtering. Al(3nm) was used as a capping layer to protect the CoFeB films and to provide an additional interface for backflow current (cf. section methods). The deposited CoFeB films were investigated by grazing incidence X-ray diffraction (GIXRD) as well as x-ray reflectivity (XRR) with Cu K$_\alpha$ radiation (cf. section methods). It was found that all CoFeB films are amorphous in nature as there is no Bragg peak for the CoFeB films as shown in Fig. 2(a). The film thickness and interface roughness were investigated by simulating the XRR spectra (red lines) using the GenX software \cite{bjorck2007genx}. The observed and simulated XRR spectra are shown in Fig. 2(b). The estimated values of film thicknesses and both interface roughnesses for our stack are presented in Table \ref{tbl:XRR and thickness}. Here, we found that the $\sigma_{CoFeB/capping}$ is larger than $\sigma_{substrate/CoFeB}$, which contributes to the net back flow current. 

\begin{table}
\small
  \caption{Comparison of nominal ($t_{CoFeB}^N$) and GenX \cite{bjorck2007genx} extracted  thickness ($t_{CoFeB}^E$) from XRR analysis. The thickness and roughness are in nm.}  \label{tbl:XRR and thickness}
  \begin{tabular*}{0.94\textwidth}{@{\extracolsep{\fill}}llllllll}
    \hline
    \hline
    & $t_{CoFeB}^N$ & & $\sigma_{substrate/CoFeB}$ & $t_{CoFeB}^E$ & $\sigma_{CoFeB/capping}$ & t$_{Al_2O_3}^E$ & $\sigma_{capping/air}$ \\
    &  & & ($\pm$0.05)& ($\pm$0.05) & ($\pm$0.05) & ($\pm$0.05) & ($\pm$0.05) \\
    \hline
    \hline
     & 2 & &0.27 & 1.89 & 0.91 & 3.49 & 1.21\\
    & 3 & &0.55& 3.26 & 0.82 & 3.46 & 1.89\\
    & 5 & & 0.69 & 5.16 & 0.77 & 3.19 & 1.99\\
    & 6 & & 0.59 & 6.13 & 0.80 & 3.20 & 1.99\\
    & 10 & & 0.28 & 11.33 & 0.87 & 3.09 & 1.19\\
    & 15 & & 0.20 & 15.95 & 1.35 & 3.95 & 1.84\\
    \hline
    \end{tabular*}
\end{table}

Figure 1 illustrates a schematic of THz emission from AHE-based CoFeB emitters when the laser pulse is irradiated from the substrate and capping sides. Due to different interface roughness (cf. Table 1), a net backflow current is expected. According to the relation $j_t = \theta_{AHE}$ ($m \times j_{bf}$), the direction of the net backflow current depends on two factors; the direction of magnetization ($m$) and the direction of the pumping side (i.e., laser pumping either from the substrate side or the capping side). 

Terahertz emission due to the iSHE can be neglected, since there is no heavy metal with large spin-orbit coupling (i.e., high S2C conversion efficiency) in our films. Instead, the origin of the recorded THz emission can result from a combination of the UDM and AHE mechanisms. Figure 3(a) shows the generated THz waveforms from a substrate/CoFeB(5nm)/cap emitter with different magnetization polarities and different pumping sides. Note that the polarity of the THz waveform is reversed when reversing the magnetization direction while keeping the pumping side the same (i.e., the net backflow current direction is the same). Thus, the THz emission is clearly of magnetic origin. Moreover, the THz waveform polarity is reversed when the sample is flipped, which is attributed to the AHE due to a change in direction of the net backflow current. It should be noted that the waveform polarities will be the same for the UDM mechanism due to similar magnetization dynamics. 
To distinguish between the contributions from the AHE and UDM to the THz emission, the approach of Liu et al.\cite{PhysRevB.104.064419} was followed by comparing the THz waveform emitted from the capping and substrate sides of the emitter. The difference between these waveforms reveals that the AHE is the dominant effect, contributing $\sim$93$\%$ of the THz emission, while UDM contributes only $\sim$7$\%$.
Thus, this clearly indicates that the dominant THz emission mechanism for the CoFeB emitters is the AHE, which is a combination of bulk and interface effects. The dependence of the THz peak-to-peak amplitude ($E_{THz}^{P-P}$) on the laser pump fluence is shown in the inset of Fig. 3(b). Note that $E_{THz}^{P-P}$ shows a linear behaviour up to 0.18 mJ/cm$^2$ pump fluence, thus  the pump fluence was fixed at 0.18 mJ/cm$^2$ for the thickness-dependent study as discussed below.

\begin{figure}
    \centering
    \includegraphics[width=12cm]{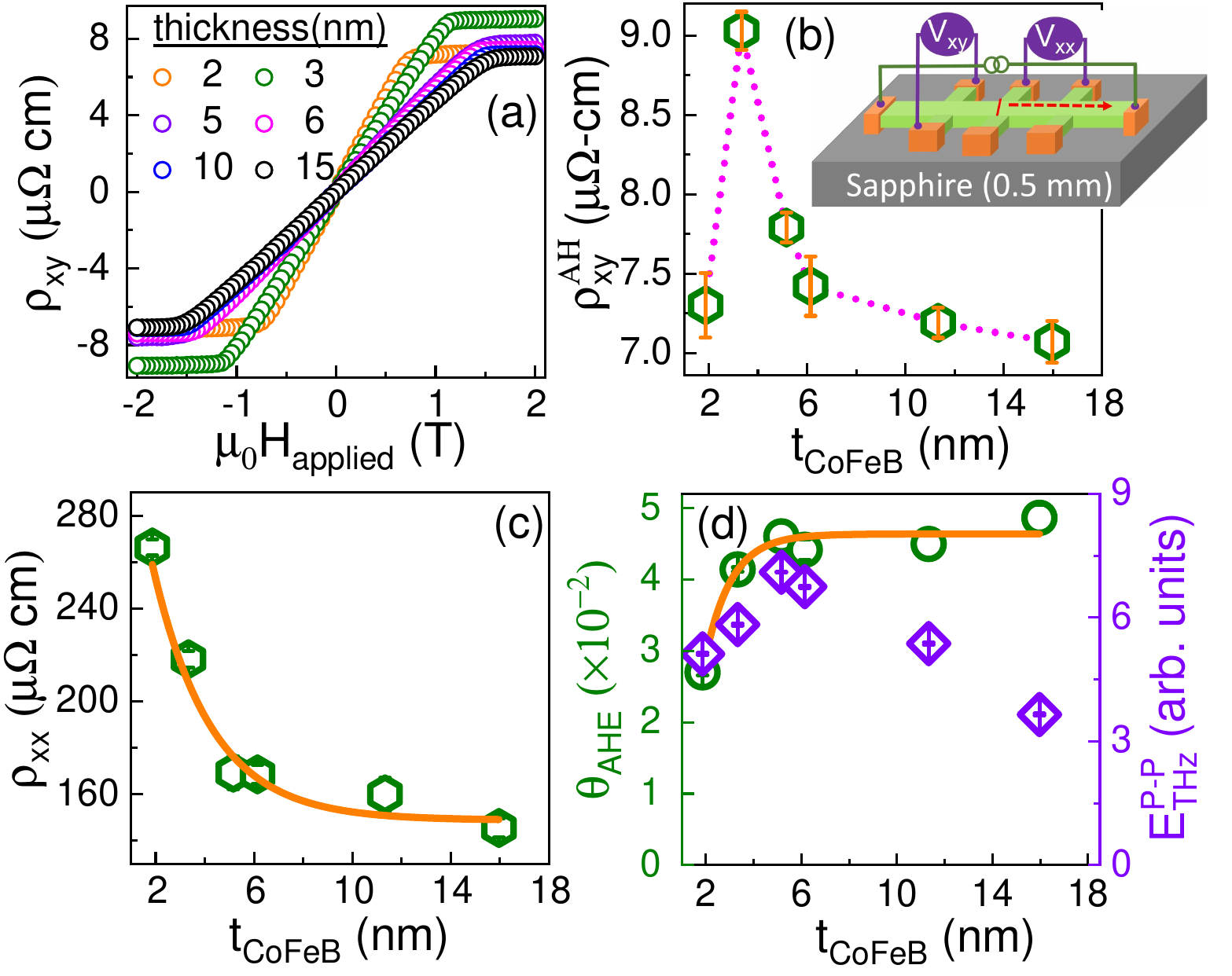}
        \caption{Results from Hall ($\rho_{xy}$) and longitudinal ($\rho_{xx}$) resistivity measurements on CoFeB films with different thickness. (a)  $\rho_{xy}$  vs. applied magnetic field ($\mu_0 H_{applied})$,   (b) $\rho_{xy}^{AH}$  vs. $t_{CoFeB}$ where $\rho_{xy}^{AH}$ corresponds to the saturation value of $\rho_{xy}$, (c) $\rho_{xx}$ vs. $t_{CoFeB}$, and (d) THz peak-to-peak amplitude (right y-axis)  and  anomalous Hall angle (left y-axis) vs. $t_{CoFeB}$.}
    \label{fig:AHE}
\end{figure}

The CoFeB thickness dependence of the AHE was measured to further investigate the relationship between the THz emission and the AHE. A schematic of the AHE measurement using a Hall bar geometry is depicted in the inset of Fig. 4(b). A longitudinal DC current (4 mA) was applied and the transverse voltage (i.e., Hall voltage) was  measured while sweeping the out-of-plane magnetic field. In Fig. 4(a), the Hall resistivity  ($\rho_{xy}$) contains contributions from the AHE and the ordinary Hall effect. The anomalous Hall resistivity ($\rho_{xy}^{AH}$), defined as the saturation value of $\rho_{xy}$, was extracted after subtracting the contribution from the ordinary Hall effect; the thickness dependence of $\rho_{xy}^{AH}$ is shown in Fig. 4(b). The CoFeB thickness dependence of the longitudinal resistivity ($\rho_{xx}$) was also measured, which follows an exponential behavior as shown in Fig. 4(c). The ratio of the anomalous Hall resistivity to the  longitudinal resistivity defines the anomalous Hall angle ($\theta_{AHE} = \rho_{xy}^{AH}/\rho_{xx}$). A comparison of the $\theta_{AHE}$ and the THz peak-to-peak amplitude as a function of CoFeB thickness is shown in Fig. 4(d). Both $\theta_{AHE}$ and $E_{THz}$ increase and follow a similar trend up to about 5 nm CoFeB thickness. However, while  $\theta_{AHE}$ saturates with a further increase of CoFeB thickness, $E_{THz}^{P-P}$ decreases. The increase of the THz emission  amplitude up to 5nm CoFeB thickness may also be explained by that the applied field ($\sim$ 85 mT) during the THz emission experiment is not sufficient to completely saturate the CoFeB films as CoFeB may possess an out-of-plane or easy cone anisotropy in the low thickness regime. Therefore, in-plane magnetization hysteresis loops were measured for all films, clearly showing that this is not a valid explanation for the studied films; the magnetization for all CoFeB films saturates at magnetic fields smaller than 85 mT as shown in the inset of Fig. 5(b). The decrease of the THz emission amplitude at larger CoFeB thickness can be explained by the fact that the spin current generated by the laser pulse scales as $A/t_{CoFeB}$ \cite{gupta2021co2feal}, where $A$ is the absorption  of the incident pump pulse by the emitter and $t_{CoFeB}$ is the thickness of the CoFeB layer. Therefore, the absorption of the incident laser pulse was calculated by measuring the incident ($P_{inc}$), reflected ($P_{ref}$) and transmitted ($P_{trans}$) powers in the optical range for all  CoFeB films. The absorption was calculated from $A = 1-T-R$, where $T=P_{trans}/P_{inc}$ and $R=P_{ref} / P_{inc}$ are the transmittance and reflectance, respectively. The absorption as a function of CoFeB thickness is shown in Fig. 5(a), which decreases from 60\% to 45\% as the CoFeB thickness increases.   Note that the change in absorption ($\sim$33\%) is not enough to explain the decrease of the THz emission ($\sim$2 times). Instead the results show that the decrease of the  THz emission amplitude is dominated by the thickness effect (i.e., the Fabry-Perot cavity effect \cite{seifert2016efficient,gupta2021co2feal}). Furthermore, the self-absorption of THz radiation also leads to a decrease of the THz emission intensity in the large thickness regime.  In our previous work \cite{gupta2021co2feal}, we reported that the self-absorption scales as $e^{-t_{CoFeB}/\zeta_{THz}}$, where $\zeta_{THz}$ is the effective inverse attenuation coefficient of the THz radiation. For metallic THz emitters, $\zeta_{THz}$ ranges from 14-22 nm \cite{gupta2021co2feal,PhysRevApplied.12.054027,torosyan2018optimized}. Therefore, the decrease of the THz emission intensity at larger thickness is a combined effect of thickness and self-absorption of THz radiation in the CoFeB-based emitter.

The experimental results confirm that the dominant mechanism for THz emission in CoFeB is the AHE. However, the AHE can have both intrinsic and extrinsic contributions. The intrinsic contribution for the AHE arises from the electronic band structure and the Berry phase \cite{RevModPhys.82.1539}, which generally requires growth of epitaxial layers. A dominant intrinsic contribution to the AHE in our case is unlikley since the CoFeB films are amorphous in nature. Moreover, the extrinsic contribution originates  from impurity scatterings such as side jump and skew scattering  \cite{RevModPhys.82.1539}. The anomalous Hall resistivity is expected to be proportional to the saturation magnetization ($M_s$) \cite{RevModPhys.82.1539}, i.e., $\rho_{xy}^{AH}=R_sM_s$, where $R_s$ is the anomalous Hall coefficient. $R_s$ increases linearly with $\rho_{xx}$ for skew scattering (i.e., $R_s$ $\propto \rho_{xx}$), while the dependence is parabolic for side jump scattering  (i.e., $R_s \propto \rho_{xx}^2$). Therefore, the CoFeB thickness dependence of $M_s$ was measured to be able to identify the dominant scattering mechanism for CoFeB; the results are shown in Fig. 5(b). Figure 5(c) shows $\rho_{xy}^{AH}/M_s$ versus  $\rho_{xx}$, indicating a linear relationship. Here, the anomalous Hall resistance (R$_s$ = $\rho_{xy}^{AH}$/M$_s$) 
is significantly affected for 2 nm nominal CoFeB thickness, deviating strongly  from the linear relationship observed at larger thickness. The deviation is enhanced by the magnetic dead layer (DL), which was calculated to be 0.59 nm (cf. Fig. 5 (d)), resulting in an increased impact of interface scattering on the charge carrier transport  in the low thickness regime. Additionally, the substrate/CoFeB interface roughness is found to be smaller than that of the  CoFeB/capping interface  (cf. Table 1). This indicates that the CoFeB/capping interface plays a dominant role for the longitudinal resistivity at low film thickness. The deviation for the 2 nm thick CoFeB emitter is thus explained by that the electrical transport at such low thickness is dominated by surface/interface  scattering and that this scattering may not be an efﬁcient mechanism for AHE since it is mostly spin-independent \cite{yang2018anomalous}. Note that the anomalous Hall angle measurement is based the diffusive transport model, while the THz emission involved the super-diffusive transport model. The conduction electrons involved in the super-diffusive transport model are located close to the Fermi level, making them equivalent to the conduction electrons in the diffusive transport model. Seifert \textit{et al.}   \cite{seifert2021frequency} demonstrated that the anomalous Hall conductivity and anomalous Hall angle are frequency-independent up to 40 THz, and therefore consistent with the DC anomalous Hall conductivity within an error bar of 2\%.

\begin{figure}
    \centering
    \includegraphics[width=10cm]{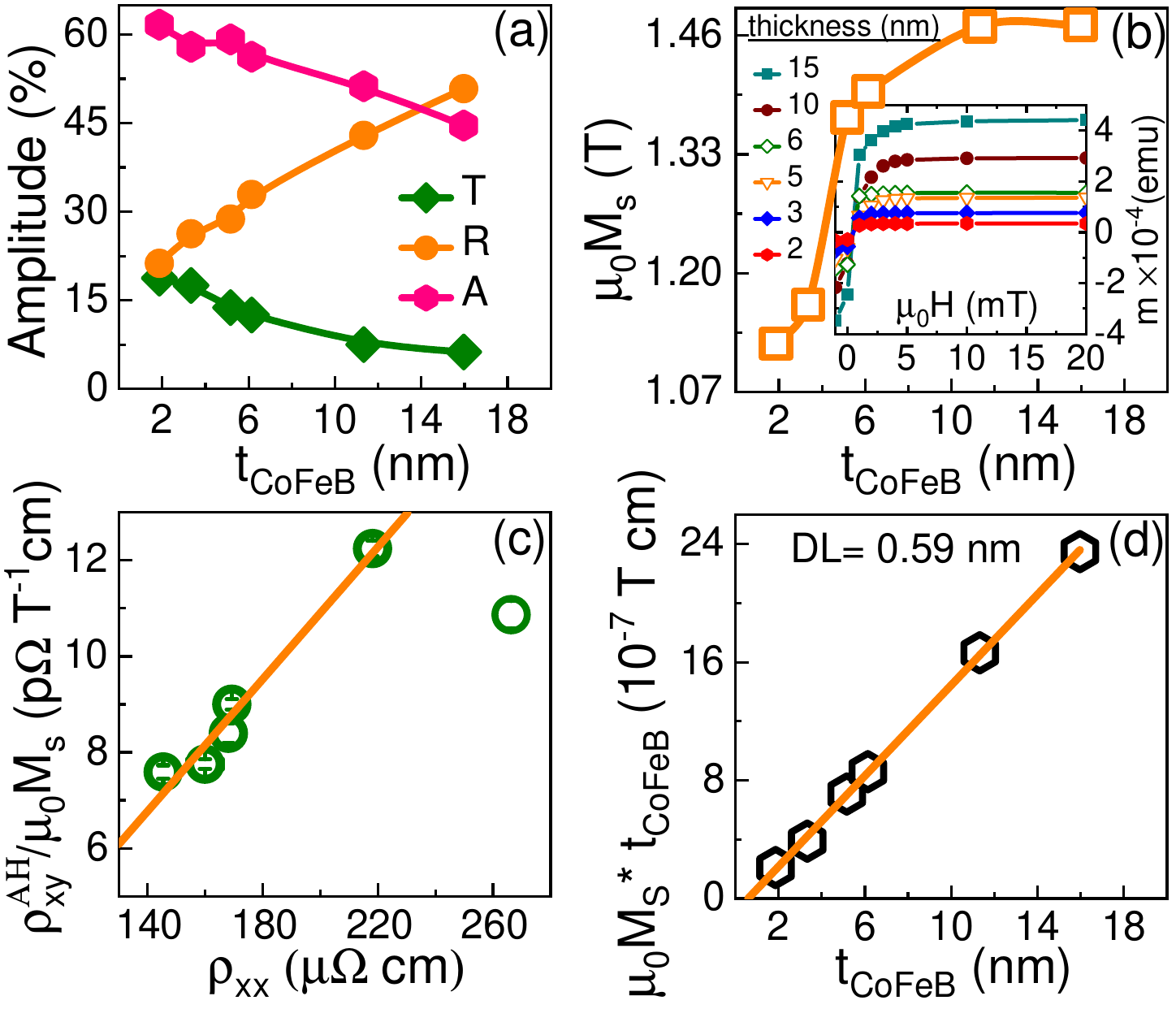}
        \caption{(a) Transmission ($T$), reflectance 
 ($R$) and absorption ($A$) vs. $t_{CoFeB}$. Solid lines are shown as guide to the eye. (b) Saturation magnetization $\mu_0M_s$ vs. $t_{CoFeB}$. Solid line is shown as guide to the eye. (c) $\rho_{xy}^{AH}/\mu_0M_s$ vs. $\rho_{xx}$. Solid line is a linear fit excluding the result for the 2 nm thick film. (d) $\mu_0M_s \cdot t_{CoFeB}$ vs. $t_{CoFeB}$. Solid line is a linear fit to the experimental data. 
 }\label{fig:my_label}
\end{figure}

\section*{Conclusions}
To summarize, we provide direct evidence that THz emission in CoFeB thin films is a consequence of the anomalous Hall effect. The THz emission from single layer CoFeB  films has been studied as a function of CoFeB thickness and the laser pump fluence. It is shown that the THz emission is a result of the nonthermal spin-polarized currents created by the laser pulse, which are reflected at the interfaces of the ferromagnetic layer, forming a net backflow current that is converted to a transverse charge current via the anomalous Hall effect. Results from electrical transport measurements show that the anomalous Hall angle increases with an increase of CoFeB layer thickness up to about 5 nm of CoFeB, after which it saturates with further increase of the thickness. This can be contrasted with the thickness dependence of the emitted THz amplitude, for which the increase up to about 5 nm is superseded by a decrease due to the Fabry-Perot cavity effect. Lastly, a detailed analysis of the results obtained from electrical transport measurements indicates that skew scattering is the dominant mechanism in our anomalous Hall effect-based CoFeB THz emitter.

\section*{Methods}

Thin films of CoFeB with different thicknesses ($t_{CoFeB}$) were deposited at room temperature using a DC magnetron sputtering system equipped with a turbo pump achieving a base pressure of 1$\times$10$^{-10}$ Torr. The base pressure and working pressure of the chamber during deposition were $5\times 10^{-10}$ and $2\times 10^{-3}$ Torr, respectively where 99.999\% pure Ar gas was used as a sputtering gas. Prior to the deposition, the single-side polished Al$_2$O$_3$ substrates (thickness $\sim$ 0.5 mm) were preheated at base pressure to 600$^{\circ}$C for one hour and then cooled down to room temperature before deposition. Substrates were rotated at 6 rpm speed for uniform growth. The deposition rate of CoFeB was calibrated using XRR measurements. To protect the CoFeB layer from oxidation and provide an additional interface for the backflow current, a 3 nm thick Al cap layer was deposited. The Al cap becomes completely oxidized at ambient conditions.

In GIXRD, the Cu K$_{\alpha}$ x-ray radiation was incident on the emitters with an incident angle of $\omega$ = 1$^{\circ}$. The GIXRD scans were performed in continuous scan mode with the scattering angle, 2$\theta$, in the range of 10$^{\circ}-$75$^{\circ}$.

A THz time-domain spectrometer was used to measure the THz emission from the CoFeB films \cite{gupta2021co2feal,gupta2021strain,gupta2021spin}. A Spectra-Physics Tsunami (Ti: sapphire) was used as a laser source, which delivers pulses of $\sim 55$ fs duration (bandwidth $\sim 12$ nm, central wavelength $\sim 800$ nm, and maximum output energy $\sim 10$ nJ) at a repetition rate of 80 MHz.  A low-temperature gallium arsenide photoconductive dipole antenna with $\sim 4 \mu$m gap was used as a detector for the THz pulses. A probe beam with average laser  power of 10 mW was used for detection and a static in-plane magnetic field of $\sim 85$ mT was used to saturate the magnetization of the CoFeB films. The displayed THz signals in Fig. 3(a) correspond to averages of 500 detected THz spectra obtained within one minute of measurement time.

The Quantum Design physical property measurement and magnetic properties measurement systems were employed to measure the AHE and the saturation magnetization of the CoFeB-based emitters, respectively. For AHE measurements, Hall bar devices (20$\times$100 $\mu$m$^2$) were patterned using optical lithography. 

\bibliography{Bibliography}

\section*{Acknowledgements}
This work is supported by the Swedish Research Council (grant numbers 2021-04658, 2018-04918, and 2017-03725) and Olle Engkvists Stiftelse (grant number 182–0365).

\section*{Author contributions statement}
R.G. conceived the idea. P.R. deposited the film and perform the XRD and XRR measurements with the support from R.G. V.M. and R.G. performed terahertz measurements with input from A.R. R.B. fabricated the Hall bar devices. V.M. and R.G. performed anomalous Hall measurements. V.M. wrote the manuscript with support from R.G. All  authors  participated in the interpretation of data. P.S. and R.G. finalized the manuscript.

\section*{Data availability}
The datasets used and/or analyzed during the current study are available from the corresponding author on reasonable request.
\end{document}